\documentstyle{article}

\date{}
\newcommand{\tr}{\,\mbox{tr}\,}

\begin{document}

\begin{center}
{\Large \bf An approach to approximate calculations of Green
functions}\\

\vspace{4mm}

 V.E. Rochev \\
 Institute for High Energy Physics \\
 Protvino, Moscow region, Russia\\
\end{center}

\begin{abstract}
A new approach proposed  recently by author for the calculation of
Green functions in quantum field theory and quantum mechanics  is
briefly reviewed. The method is applied to nonperturbative
calculations for anharmonic oscillator, $\phi^4$-theory, quantum
electrodynamics and other models.

\end{abstract}

\section{Toy model}

To illustrate the general properties of the calculation scheme,
which was proposed recently in works \cite{Ro1}-\cite{RoSap1}, we
consider a toy problem of the calculation of "$n$-particle Green
functions" in zero-dimensional theory with $\phi^4$-interaction,
i.e., a problem of calculating the following quantities
\begin{equation}
G_n=g_n/g_0,
\end{equation}
 where
\begin{equation}
g_n=\int^\infty_{-\infty} d\phi\, \phi^{2n}\exp
\{-\frac{m^2}{2}\phi^2-\lambda\phi^4\}.
\end{equation}
A generating function for these quantities is
\begin{equation}
g(x)= \int^\infty_{-\infty} d\phi\exp
\{-\frac{m^2}{2}\phi^2-\lambda\phi^4+x\phi^2\}.
\end{equation}
At that $g_n=\frac{d^ng}{dx^n}\mid_{x=0}$.

The generating function $g(x)$ satisfies the differential equation
\begin{equation}
4\lambda\frac{d^2g}{dx^2}+(m^2-2x)\frac{dg}{dx}-g=0,
\end{equation}
which is the Schwinger-Dyson equation (SDE) for this toy model.

A solution of this equation can be written as a coupling constant
perturbation (CCP) series $ g(x)=g^{(0)}_{pert}+\cdots+
g^{(i)}_{pert}+\cdots$ where leading ("free field") approximation
is $g^{(0)}_{pert}= \exp\{-\frac{1}{2}\log|m^2-2x|\}$ (up to a
multiplier which inessential for calculating of $G^{(i)}$)   and
$g^{(i)}_{pert}$ is a solution of the perturbation iteration
scheme equation
$$(m^2-2x)\frac{dg^{(i)}_{pert}}{dx}-g^{(i)}_{pert}
=-4\lambda\frac{d^2g^{(i-1)}_{pert}}{dx^2}.$$

From the point of view of differential equation theory the CCP is
attributed to the type of so-called singular perturbations. This
fact defines, to a considerable extent, both the poor convergence
properties (the CCP series is an asymptotic expansion at best) and
the limited nature of the field of its applicability. (In the
model under consideration the CCP theory gives good results in the
region $\lambda\leq 0.01m^4$ and nothing more.) A perturbation is
named to be singular if it contains a higher derivative term. The
perturbation theory over $\lambda$ is singular  in the above
sense, since the leading approximation consists in the neglecting
of the higher derivative term. (See \cite{RoSap2}, \cite{Ro2} for
more discussion.)

An alternative for the perturbation theory over $\lambda$ can be
other iterative scheme that based on an approximation of eq. (1)
near the point $x=0$ by an equation with constant coefficients.
Take as a leading approximation the equation
\begin{equation}
4\lambda\frac{d^2g^{(0)}}{dx^2}+m^2\frac{dg^{(0)}}{dx}-g^{(0)}=0.
\end{equation}

 The term $2x\frac{dg}{dx}$ will be considered as a perturbation.
The iteration scheme will  consist in step-to step solutions of
inhomogeneous equations
\begin{equation}
4\lambda\frac{d^2g^{(i)}}{dx^2}+m^2\frac{dg^{(i)}}{dx}-g^{(i)}=
2x\frac{dg^{(i-1)}}{dx}.
\end{equation}
  Leading approximation equation has the solution
  $g^{(0)} = e^{\alpha x}$, where $\alpha$'s are roots of the
  characteristic equation
$$4\lambda\alpha^2+m^2\alpha-1.$$ From a condition of the
convergence of integral (3) at $\lambda\rightarrow 0$ we choose
the root $$ \alpha=\frac{-m^2+ \sqrt{m^4+16\lambda}}{8\lambda}=
1/m^2 + {\cal O}(\lambda). $$ A solution of $n$th-step equation is
looked for in the form $g^{(i)}=p^{(i)}g^{(0)}$. Taking into
account the leading approximation equation (5), we obtain the
equation for $p^{(i)}$
\begin{equation}
4\lambda\Big(\frac{d^2p^{(i)}}{dx^2}+2\alpha
\frac{dp^{(i)}}{dx}\Big)+m^2\frac{dp^{(i)}}{dx}=2x\Big(\frac{dp^{(i-1)}}{dx}
+\alpha p^{(i-1)}\Big).
\end{equation}
It is evident from this equation that $p^{(i)}$ is a polynomial of
degree $2i$ in $x$.

   The question about a
small parameter for the expansion defined by eqs. (5)-(7) arises.
There is no  manifest small parameter for this expansion , but it
is clear, that the expansion approximates well the exact solution
not only for small values of $\lambda$. Really, in the strong
coupling region $\lambda\rightarrow\infty$ the first-step
one-particle function $G_1$ ("propagator") approximates the exact
result with accuracy of 26\%, and second-step one with accuracy
7\%.

 To be more exact, the
question about the small parameter should be replaced by the
question about a convergence of the expansion. But the convergence
of this iteration scheme can be easily proved. Notice, the
iteration scheme  defined by eqs. (5)-(7) is equivalent to
iterations of the second kind Volterra equation with the
continuous kernel.
 The convergence of the iteration of this equation is fulfilled by the
textbook theorem.

So, this regular expansion possesses  the good convergency
properties in the framework of this simple zero-dimensional model
and, more importantly, is a nonperturbative method of calculations
of Green functions.

\section{$\phi^4$-theory}

   Let us go to the field theory.
Consider the  theory of a scalar field $\phi(x)$ in the Euclidean
space $ E_d$ with the action
\begin{equation}
A(\phi) = \int dx \{\frac{1}{2} (\partial_{\mu}\phi)^2
 +\frac{m^2}{2} \phi^2 + \lambda\phi^4\}
\end{equation}and with the generating functional of Green functions (vacuum
expectation values)
\begin{equation}
G(\eta) =  \int D\phi \exp \{ -A + \phi\eta\phi \}.
\end{equation}Here $\eta(x,y)$ is a bilocal source,
$\phi\eta\phi\equiv\int dxdy\phi(x)\eta(x,y)\phi(y)$. A
normalization constant is omitted.
 The $n$th
derivative of $G$ over $\eta$ with the source being switched off
is the $2n$-point ($n$-particle) Green function.

The Schwinger-Dyson equation (SDE) for the generating functional
$G(\eta)$ is a corollary of the identity
\begin{equation}
0 = \int D\phi \frac{\delta}{\delta\phi (x)}(\phi(y)
 \exp \{ -A + \phi\eta\phi \}).
\end{equation}
Taking into account the above definitions we get  the SDE  for
$\phi^4$-theory
\begin{equation}
4\lambda \frac{\delta^2 G}{\delta\eta(y,x)\delta\eta(x,x)} + ( m^2
- \partial\,^2)\frac{\delta G}{\delta\eta(y,x)} - 2\int
\eta(x,u)\frac{\delta G}{\delta\eta(y,u)}du - \delta(x-y) G = 0.
\end{equation}
At $d=0$ (zero-dimensional theory, or "single-mode approximation")
the functional derivatives transform into usual ones, and eq.
(11), after obvious redesignations, reduces to the ordinary
differential eq. (4). At $d=1$ the model corresponds to the
quantum-mechanical anharmonic oscillator. At $d \ge 2$ (field
theory) for the cancellation of ultraviolet divergences the
appropriate counterterms should be included in the action. The SDE
for the theory with counterterms has the form of eq. (11) with the
substitution $ \lambda\rightarrow \lambda + \delta\lambda,\,\,\,\,
m^2\rightarrow m^2 + \delta m^2,\,\,\,\,\partial^2\rightarrow
(1+\delta z)
\partial^2,$
 where $\delta\lambda,\delta m^2$ and $\delta z$ are
correspondingly counterterms of coupling, mass and wave function
renormalizations.

Let apply to  SDE (11) the same idea about the approximation by an
equation with "constant" (i.e., independent from $\eta$)
coefficients.
 As the leading  approximation equation we will  consider the equation
\begin{equation}
4\lambda \frac{\delta^2 G^{(0)}}{\delta\eta\delta\eta} + ( m^2 -
\partial\,^2)\frac{\delta G^{(0)}}{\delta\eta}
 -  G^{(0)} = 0,
\end{equation}and the term $2\eta\frac{\delta G}{\delta\eta}$ (that contains
the source $\eta$ explicitly) should be treated as a perturbation.
Since Green functions are the derivatives of $G(\eta)$ in zero and
the only behaviour of $G$ near $\eta = 0$ is essential , such an
approximation seems to be acceptable. The iteration procedure for
the generating functional $G = G^{(0)} + G^{(1)} + \cdots +
G^{(i)} + \cdots$ consists in the step-to-step solution of the
equations
\begin{equation}
4\lambda \frac{\delta^2 G^{(i)}}{\delta\eta\delta\eta} + ( m^2 -
\partial\,^2)\frac{\delta G^{(i)}}{\delta\eta}
 -  G^{(i)} =
2\eta\frac{\delta G^{(i-1)}}{\delta\eta}. \end{equation}The
solution of the leading approximation equation (10) is the
functional
\begin{equation}
G^{(0)} = \exp \{\int dx dy \eta(y,x)\bigtriangleup^{(0)}(x-y)\},
\end{equation}where $\bigtriangleup^{(0)}$ is a solution of the 
"characteristic" equation
\begin{equation}
4\lambda\bigtriangleup^{(0)}(0)\bigtriangleup^{(0)}(x-y) + ( m^2 -
\partial\,^2)\bigtriangleup^{(0)}(x-y) = \delta(x-y). \end{equation}
 At $d\geq2$ the
quantity $\bigtriangleup^{(0)}(0)$ must be considered as some
regularization.

Equation~(15) is  similar to the equation for the propagator in
the leading approximation of the $1/N$-expansion. Certainly, the
similarity is completely superficial, since the principle of the
construction of the approximation scheme is
 different.

The solution of equation (15) is the free propagator
$\bigtriangleup^{(0)} = ( \mu^2 - \partial\,^2)^{-1}$ with the
renormalized mass $\mu^2=m^2+4\lambda\bigtriangleup^{(0)}(0)$. The
quantity $\bigtriangleup^{(0)}(0)$ is defined from the
self-consistency condition. The propagator is the first derivative
of $G(\eta)$ over the source $\eta :\,\,\,
G_1\equiv\bigtriangleup={\frac{\delta G}{\delta\eta}}\mid_{\eta\,
=0}$. As can be easily  seen, it is simply $\bigtriangleup^{(0)}$
for the leading approximation.

Notice, that the generating functional (14) of leading
approximation does not possess the complete Bose-symmetry. Really,
as follows from the definition of generating functional, the
Bose-symmetry implies on {\it full} generating functional the
condition
\begin{equation}
\frac{\delta^2G}{\delta\eta(y,x) \delta\eta(y',x')}=
\frac{\delta^2G}{\delta\eta(y',x) \delta\eta(y,x')}. \label{bose}
\end{equation} Evidently condition (\ref{bose}) does not
fulfilled for $G^{(0)}$ defined by eq.(14). The violation of this
condition leads particularly to the violation of connected
structure of the leading approximation two-particle (four-point)
Green function.

Such a situation is rather typical for nonperturbative
calculational schemes with  bilocal source (for example, for
$1/N$-expansion in the bilocal source formalism), but discrepancy
of  such type are not an obstacle for using
 these iteration schemes. Really, condition
(\ref{bose}) should be satisfied by the {\it full} generating
functional $G$ which is an {\it exact} solution of SDE.
 It is clear that an approximate
solution may do not possess all properties of an exact one. In
given case we have  just the same situation. Properties of
connectivity and Bose-symmetry of higher Green functions, which
are not fulfilled at first steps of the iteration scheme,
"improves" at subsequent steps. For example, the structure of
disconnected part of the two-particle function is reconstructed as
early as at the first step of the iteration scheme. At subsequent
steps the correct connected structure of many-electron functions
and other corollaries of Bose-symmetry are reconstructed. Such
stepwise reconstruction of exact solution properties is very
natural for the given iteration scheme as  it is based on an
approximation of the generating functional $G(\eta)$ in vicinity
of zero. The Green functions are coefficients of the generating
functional expansion in the vicinity of zero, therefore only the
lowest functions are well-described at first steps of the
approximation -- at the leading approximation the  propagator
only. Higher many-particle functions come into the play later, at
following steps, and relation  (\ref{bose}) is fulfilled more and
more exactly when we go toward exact solution.

In the general case, the solution of  equation  for the $i$-th
step
 of the iteration scheme is the functional $G^{(i)} = P^{(i)} G^{(0)}$,
where $P^{(i)}$ is a polynomial in $\eta$ of a degree $2i$.
Therefore at the $i$-th step the computation of Green functions
reduces to
 solving  a system of $2i$ linear integral equations.

A solution of the first step equation is $G^{(1)}=P^{(1)}G^{(0)}$
where $P^{(1)}=\frac{1}{2}F\eta^2+\bigtriangleup^{(1)}\eta$. The
function $F$ is two-particle (four-point) function of the  first
step, and $\bigtriangleup^{(1)}$ is the first-step correction to
the propagator.  Eq.(13) at $i=1$
 gives us a system of equations for
$F$ and $\bigtriangleup^{(1)}$. Equations for $F$ and
$\bigtriangleup^{(1)}$ are simple linear integral equations. The
exact form of solutions of these equations see in~\cite{Ro1}. At
$\lambda \rightarrow 0$ the first step propagator reproduces
correctly the first term of the usual CCP theory.

At  $d=1$ the model  with action (6) describes the
quantum-mechanical anharmonic oscillator. Ultraviolet divergences
are absent,  quantities of $\bigtriangleup^{(0)}(0)$ type
 are finite and the above formulae  are applied
directly for the computation of Green functions.

To calculate a ground state energy $E$ one can use the well-known
formula $$ \frac{dE}{d\lambda} = G_2 (0,0,0,0),$$ where $G_2$ is
the four-point (or two-particle) function. Integrating the formula
with  a boundary condition $E(\lambda=0)=m/2$ taken into account,
one can calculate the ground state energy for all values of the
coupling (see \cite{Ro1}).

At  $\lambda\rightarrow 0$ the first step calculation
  reproduces the perturbation theory up to the second order.
At $\lambda\rightarrow\infty:\,\,\,E=\epsilon_0\lambda^{1/3} +
O(\lambda^{-1/3})$, and $\epsilon_0 = 0.756$. The coefficient
$\epsilon_0$ differs by 13\% from the exact numerical one
$\epsilon_0^{exact} = 0.668$. At $\lambda/m^3 = 0.1$ the result of
the calculation differs from the exact numerical one by 0.8\% and
at $\lambda/m^3 = 1$ differs by 6.3\%. Therefore, the first step
calculations approximate the ground state energy for all values of
$\lambda$ with the accuracy that varies smoothly from 0 (at
$\lambda\rightarrow 0$) to 13\% (at $\lambda\rightarrow\infty$).

At $d\geq 2$  action (6) should be added by  counterterms for the
elimination of ultraviolet divergences. There is no need to add a
counterterm of wave function renormalization for the leading
approximation, and the equation of the leading approximation will
be
\begin{equation}
4(\lambda+\delta\lambda_0) \frac{\delta^2
G^{(0)}}{\delta\eta\delta\eta} + (\delta m^2_0 +  m^2 -
\partial\,^2)\frac{\delta G^{(0)}}{\delta\eta}
 -  G^{(0)} = 0.
\end{equation}At $i\geq 1$ the counterterms $\delta\lambda_i$,
 $\delta m^2_i$
and $\delta z_i$ should be considered as perturbations. Therefore,
the corresponding terms should be added to the r.h.s. of  equation
(11). So, the first step equation will be \begin{eqnarray}
4(\lambda+\delta\lambda_0) \frac{\delta^2
G^{(1)}}{\delta\eta\delta\eta} + (\delta m^2_0 +  m^2 -
\partial\,^2)\frac{\delta G^{(1)}}{\delta\eta}
 -  G^{(1)} =\nonumber\\
= 2\eta\frac{\delta G^{(0)}}{\delta\eta} - \delta
m^2_1\frac{\delta G^{(0)}}{\delta\eta} + \delta z_1
\partial^2\frac{\delta G^{(0)}}{\delta\eta} - 4\delta\lambda_1
\frac{\delta^2 G^{(0)}}{\delta\eta\delta\eta}. \end{eqnarray}

For the super-renormalizable theory ($d=2$ and $d=3$) it is
sufficient to add counterterms of mass renormalization  and wave
function renormalization, i.e. $\delta\lambda_i=0$ for all $n$.
The normalization condition on the physical renormalized mass
$\mu^2$ gives us  a counterterm of  the mass renormalization in
the leading approximation. This counterterm diverges
logarithmically at $d=2$ and linearly at $d=3$. The counterterm
$\delta z_1$ is finite at $d=2,3$. The counterterm $\delta m^2_1$
diverges as that of the leading approximation does,
 namely, logarithmically at $d=2$ and linearly at $d=3$.

At  $d=4$ besides the renormalizations of the mass and the wave
function a coupling  renormalization is necessary. Due to the
presence of the counterterm $\delta\lambda$ the normalization
condition on the renormalized mass $\mu^2$ for the leading
approximation becomes the connection between counterterms $\delta
m^2_0$ and $\delta\lambda_0$. Counterterm $\delta\lambda_0$ (and,
consequently,
 $\delta m^2_0$) will be fixed at the {\it following} step
of the iteration scheme.

A solution of the equation for the four-point function $F$ at
$d=4$ diverges logarithmically, and a renormalization of the
coupling is necessary. The equation for $F$ contains the
counterterm $\delta\lambda_0$ only. Therefore by defining a
renormalized coupling $\lambda_r$ as a value of the amplitude in a
normalization point we obtain the counterterm of the coupling
renormalization $\delta\lambda_0$ and the renormalized amplitude.
Taking the renormalization of the two-particle amplitude in  such
a manner, one can  solve the equation for $\bigtriangleup^{(1)}$
and renormalize the mass operator in correspondence with the
general principle of normalization on the physical mass. But in
four-dimensional case one gets an essential obstacle. At the
regularization removing, $\delta\lambda_0\rightarrow -\lambda$,
and the coefficient $\lambda+\delta\lambda_0$ in the leading
approximation equation (17) vanishes. The same is true for  all
the subsequent iterations. The theory is trivialized. One can
object that an expression
\begin{equation}
(\lambda+\delta\lambda_0)\cdot \frac{\delta^2
G}{\delta\eta(y,x)\delta\eta(x,x)} \end{equation}is really an
indefinite quantity  of  $0\cdot\infty$ type,
 and the renormalization is,
in the essence, a definition of the quantity. But it does not save
a situation {\it in this case} since the renormalized amplitude
possesses a nonphysical singularity in a deep-euclidean region (it
is a well-known Landau pole). The unique noncontradictory
possibility is a choice $\lambda_r\rightarrow 0$ at the
regularization removed. This is the triviality of the theory
again. The triviality appears almost inevitably in an
investigation of $\phi^4_4$-theory beyond  the CCP theory and is a
practically rigorous result. Notice, that  contrary to the CCP
theory which is absolutely nonsensitive to the triviality of the
theory, the method proposed leads to the triviality already at the
first step.

\section{Quantum Electrodynamics}

 The Lagrangian of Quantum Electrodynamics (QED) in Minkowski space-time
 with a gauge fixing
term has the form
\begin{equation}
{\cal L} = -\frac{1}{4}F_{\mu\nu}F_{\mu\nu}
-\frac{1}{2d_l}(\partial_\mu A_\mu)^2 + \bar\psi(i\hat\partial - m
+ e\hat A)\psi. \label{lagrangian} \end{equation}

Here $F_{\mu\nu}=\partial_\mu A_\nu - \partial_\nu A_\mu,\; \hat A
\equiv A_\mu\gamma_\mu,\; \bar\psi = \psi^*\gamma_0,\; m$ is an
electron mass, $e$ is a charge (coupling constant), $d_l$ is a
gauge parameter, $\gamma_\mu$ are Dirac matrices.  For notation
simplicity we write all vector indices as low ones.

A generating functional of Green functions is
\begin{equation}
G(J,\eta) = \int D(\psi,\bar\psi,A)\exp i\Big\{\int dx\Big({\cal
L} +J_\mu(x)A_\mu(x)\Big) -\int dx dy
\bar\psi^\beta(y)\eta^{\beta\alpha}(y,x)\psi^\alpha(x)\Big\}.
\label{G} \end{equation}

Here $J_\mu(x)$ is a source of the gauge field, and
$\eta^{\beta\alpha}(y,x)$ is a bilocal source of the spinor field
($\alpha$ and $\beta$ are spinor indices). Normalization constant
omitted.

Functional derivatives of $G$ with respect to sources are vacuum
expectation values
\begin{equation}
\frac{\delta G}{\delta J_\mu(x)} = i<0\mid A_\mu(x)\mid 0>,\;\;
\frac{\delta G}{\delta\eta^{\beta\alpha}(y,x)} = i<0\mid
T\Big\{\psi^\alpha(x)\bar\psi^\beta(y)\Big\}\mid 0>. \label{VEV}
\end{equation}
 SDEs for the generating functional of Green
functions of QED are
\begin{equation}
(g_{\mu\nu}\partial^2-\partial_\mu\partial_\nu+
\frac{1}{d_l}\partial_\mu\partial_\nu) \frac{1}{i}\frac{\delta
G}{\delta J_\nu(x)} + ie\, \mbox{tr}\Big\{ \gamma_\mu\frac{\delta
G}{\delta\eta(x,x)}\Big\} + J_\mu(x)G = 0, \label{SDEA}
\end{equation}
\begin{equation}
\delta(x-y)G + (i\hat\partial - m)\frac{\delta G}{\delta\eta(y,x)}
+ \frac{e}{i}\gamma_\mu\frac{\delta^2G}{\delta J_\mu(x)
\delta\eta(y,x)} - \int dx' \eta(x,x') \frac{\delta
G}{\delta\eta(y,x')} = 0. \label{SDEpsi}
\end{equation}
 (Here and everywhere
below $\partial_\mu$ denote a differentiation with respect to
variable $x$.) Let us resolve SDE
  (\ref{SDEA}) with regard to the first derivative of the generating
  functional with respect  $J_\mu$:
  \begin{equation}
  \frac{1}{i}\frac{\delta G}{\delta J_\mu(x)}=
  -\int dx_1 D^c_{\mu\nu}(x-x_1)\Big\{J_\nu(x_1)G+
  ie\tr \gamma_\nu\frac{\delta G}{\delta\eta(x_1,x_1)}\Big\},
  \label{dG/dJ}
  \end{equation}
and substitute it into the second SDE (\ref{SDEpsi}). As a result
we obtain the "integrated over  $A_\mu$" (in the
functional-integral terminology) equation
\begin{eqnarray}
 \delta(x-y)G +
(i\hat\partial - m)\frac{\delta G}{\delta\eta(y,x)} +
\frac{e^2}{i} \int dx_1 D^c_{\mu\nu}(x-x_1)\gamma_\mu
\frac{\delta}{\delta\eta(y,x)}\tr \gamma_\nu\frac{\delta G}
{\delta\eta(x_1,x_1)}= \nonumber\\ = \int dx_1 \Big\{\eta(x,x_1)
\frac{\delta G}{\delta\eta(y,x_1)} + e
D^c_{\mu\nu}(x-x_1)J_\nu(x_1)\gamma_\mu \frac{\delta
G}{\delta\eta(y,x)}\Big\}. \label{SDE}
 \end{eqnarray}
  Exploiting
Fermi-symmetry condition
\begin{equation}
\frac{\delta^2G}{\delta\eta^{\beta\alpha}(y,x)
\delta\eta^{\beta'\alpha'}(y',x')}=
-\frac{\delta^2G}{\delta\eta^{\beta'\alpha}(y',x)
\delta\eta^{\beta\alpha'}(y,x')}. \label{fermi}
\end{equation}
 let us transform
eq.(\ref{SDE}) in following manner:
 \begin{eqnarray}
\delta(x-y)G + (i\hat\partial - m)\frac{\delta G}{\delta\eta(y,x)}
+ ie^2 \int dx_1 D^c_{\mu\nu}(x-x_1)\gamma_\mu
\frac{\delta}{\delta\eta(x_1,x)}\gamma_\nu\frac{\delta G}
{\delta\eta(y,x_1)}= \nonumber\\ = \int dx_1 \Big\{\eta(x,x_1)
\frac{\delta G}{\delta\eta(y,x_1)} + e
D^c_{\mu\nu}(x-x_1)J_\nu(x_1)\gamma_\mu \frac{\delta
G}{\delta\eta(y,x)}\Big\}. \label{SDEF}
 \end{eqnarray}
 From the point of view
of {\it exact} solutions equations (\ref{SDE}) and (\ref{SDEF})
are fully equivalent since the transition from eq.(\ref{SDE}) to
eq.(\ref{SDEF}) is, in essence, an identical transformation.
However, it is not the case for the used iteration scheme since,
as for  Bose-symmetry condition (\ref{bose}) as, Fermi-symmetry
condition (\ref{fermi}) is fulfilled only approximately at any
finite step of the iteration scheme. Therefore, eqs. (\ref{SDE})
and (\ref{SDEF}) lead to different expansions. Eq.(\ref{SDE})
gives to the calculational scheme, which on the language of
Feynman diagrams of perturbation theory is analog of the summation
of chain diagrams with fermion loop. This version is named
"calculations over perturbative vacuum" since a unique connected
Green function of the leading (vacuum) approximation is the free
electron propagator. This scheme leads, as  for $\phi^4$-theory as
(see above, section 2), to Landau pole and triviality at the first
step of the iteration scheme (see \cite{Ro3} for more details).

The second version of the iteration scheme, which is based on
eq.(\ref{SDEF}), gives us a fruitful and "insensitive to
triviality" scheme of calculation of physical quantities.
 This version is named
"calculations over nonperturbative vacuum" since the electron
propagator of the leading vacuum approximation is a solution of a
non-trivial nonlinear equation. For this scheme a calculation of
 two first terms of expansion
of the vertex function in photon momentum for chiral-symmetric
vacuum have been performed in work \cite{Ro3}.  This calculation
has allowed  to obtain a simple formula for anomalous magnetic
moment:
 $f_2=\alpha/(2\pi-\alpha)$,
where $\alpha$ is the fine structure constant.
 Also, for
a linearized version of the theory (see, for instance, \cite{Mir})
the problem of dynamical chiral symmetry breaking have been
investigated in work \cite{Ro3}.
 The
calculations are performed for renormalized theory in Minkowski
space. In the  strong coupling region $\alpha\ge\pi/3$ the results
correspond to earlier investigations performed in Euclidean theory
with cutoff (see \cite{Mir}):  solutions arise
 with breakdown of chiral symmetry.
But for the renormalized theory a solution with breakdown of
chiral symmetry is also possible in the weak coupling region
    $\alpha<\pi/3$  with a subsidiary condition
    on the value of $\alpha$
    which follows from the gauge invariance (see \cite{Ro3} for more
    details).

\section{Other models}

Some other models have been investigated by proposed method in
works \cite{Ro2}-\cite{RoSap1} and \cite{Ro4}-\cite{Ro6}. Famous
Gross-Neveu model with the  Lagrangian
\begin{equation}
{\cal L} = \bar\psi_j i\hat\partial \psi_j + \frac{\lambda}{2N} (
\bar\psi_j \psi_j)^2 \end{equation} have been investigated by this
method in work \cite{RoSap1} at $D=2,3,4$ and finite $N$, where
$D$ is space dimension and $N$ is a number of flavors. The results
were following: a spontaneous symmetry breaking is shown to exist
in $D=2,3$ and the running coupling constant is calculated. The
four dimensional theory turns seems be trivial.  These results
exhibit the efficiency of the method and are the finite $N$
generalization of the known results  obtained in the framework of
$1/N$ expansion.

The greatest interest from the  physical point of view
 presents applications of the method to study gauge theories in
 nonperturbative region. First steps in this direction were made
 in works \cite{Ro4}-\cite{Ro6}. In work \cite{Ro4} a generalization
  of the Higgs mechanism  which takes
into account the contributions of gauge field vacuum configuration
into the formation of the physical vacuum was considered.
 For the Abelian Higgs model the triviality bound
 $m_H\le~1.15m_A$ was found. In works \cite{Ro5}-\cite{Ro6} a
 non-Abelian $SU(2)$ gauge theory was considered, and a mechanism for
 the dynamical mass generation of a
non-Abelian gauge field which was based on taking into account the
contributions of the gauge field vacuum configurations into the
formation of the physical vacuum was proposed. These
investigations are needed in a following elaboration. A most
winning field of application of proposed method seems to be a
problem of dynamical chiral symmetry breaking in gauge theories.

\end{document}